\documentclass[
twocolumn,
amsmath,
amssymb,
aps,
pra,
]{revtex4} % revtex4-2

\usepackage{graphicx}% Include figure files
\usepackage{dcolumn}% Align table columns on decimal point
\usepackage{bm}% bold math
\usepackage{verbatim}% Ignore the latex command
\usepackage[caption=false]{subfig}% Left-align the caption.
\usepackage{CJK}% Chinese characters can be used
\usepackage{color}

\definecolor{myColor}{rgb}{1,0,0}
\begin{document}
\begin{CJK*}{UTF8}{}
\preprint{APS/123-QED}

\title{Spatiotemporal scales of dynamical quantum phase transitions in the Bose-Hubbard model}% Force line breaks with \\

\author{Jia Li%
\CJKfamily{gbsn}(李佳)}
\affiliation{Department of Physics and Institute of Theoretical Physics, University of Science and Technology Beijing, Beijing 100083, P.~R.~China}%Lines break automatically or can be forced with \\
\author{Yajiang Hao%
\CJKfamily{gbsn}(郝亚江)}
\email{haoyj@ustb.edu.cn}
\affiliation{Department of Physics and Institute of Theoretical Physics, University of Science and Technology Beijing, Beijing 100083, P.~R.~China}
\date{\today}% It is always \today, today, but any date may be explicitly specified

\begin{abstract}
We investigate the spatial and temporal scales of dynamical quantum phase transitions in the one-dimensional Bose-Hubbard model in the strong interaction limit. Using Jordan-Wigner transformation, we obtain the time-dependent wavefunction and therefore the subsystem Loschmidt echo, and systematically investigate how its properties vary with subsystem size. It is found that when the subsystem is sufficiently large, it exhibits logarithmic divergence identical to that of the full system Loschmidt echo, yielding a critical exponent of zero. We also obtain the required subsystem size and temporal resolution for detecting dynamical quantum phase transitions using the subsystem Loschmidt echo. It is expected that the present results provide a reliable foundation for further experimental investigations.
\end{abstract}

%\keywords{None}% Use showkeys class option if keyword display desired

\maketitle
\end{CJK*}

\section{\label{sec:level1}INTRODUCTION}

Phase transitions, as one of the most remarkable phenomena in many-body systems, are characterized by the nonanalytic behavior of thermodynamic quantities at critical points, from which critical exponents satisfying scaling laws can be derived. At zero temperature, equilibrium quantum phase transitions follow a similar paradigm, classifying quantum states by calculating the variation of an observable with system parameters. For instance, in the Bose-Hubbard model \cite{Arc2016, Eji2011}, plotting the energy gap of the ground state against varying interaction strengths ($U/J$) yields a graceful phase diagram with clear boundaries---where a zero energy gap indicates a superfluid (SF) state, and a non-zero gap corresponds to a Mott insulator (MI) state. In contrast to equilibrium phase transitions, phase transitions arising in nonequilibrium processes are referred to as dynamical quantum phase transitions (DQPTs). With the rapid advancement of ultracold atomic gas experimental techniques \cite{Gro2021, Bro2020, Kja2020, Jep2020, Su2023, Fos2025}, the study of dynamical behavior in quantum many-body systems under nonequilibrium conditions has become a frontier topic in condensed matter physics and quantum information science. When a system is driven rapidly through a quantum critical point, the adiabatic theorem breaks down \cite{Sch2006, Gar2017}, preventing it from following the instantaneous ground state and giving rise to a rich spectrum of phenomena.

Fischer et al. demonstrated that in quenched two-dimensional spin-boson systems, the variance of the winding number exhibits a distinct scaling law originating from the slow decay of magnetization correlations \cite{Uhl2007}. Meanwhile, under quenches with exponentially decaying tunneling rates in the Bose-Hubbard model, the freezing of density fluctuations alongside growing phase fluctuations leads to exponentially decaying off-diagonal long-range order and a corresponding reduction in the superfluid fraction \cite{Fis2008, Sch2006}. Against this backdrop, dynamical quantum phase transitions (DQPTs) have emerged as a key paradigm, characterized by nonanalytic behavior manifested in the real-time evolution of quantum systems following a quench \cite{Blo2008, Geo2014}. Heyl et al. \cite{Hey2013, Hey2015} identified these transitions through singularities in the full system Loschmidt echo (LE)---a dynamical analogue of free energy density that quantifies the overlap between initial and time-evolved states \cite{Lah2019, Liu2019, Cao2020, Zho2021, Zen2023, Cao2023, Kul2023, Sac2024}. Their work systematically investigated the dynamical evolution of the LE and established its scaling form and critical exponents. Together with subsequent studies, they demonstrated a critical exponent of 1 in the transverse-field Ising model and a critical exponent of $1/2$ in the Ising model with both transverse and longitudinal fields \cite{Wan2025}.

While the LE offers a novel perspective for defining DQPTs, its observation poses significant experimental challenges as it requires accessing global many-body wavefunction overlap information. This has motivated the research community to pursue local or quasilocal observables capable of accurately capturing these nonequilibrium critical phenomena. Recent studies suggest that certain string operators or correlation functions may exhibit nonanalytic features at critical times \cite{Kar2025, Ban2021}, providing viable detection pathways. However, key questions remain unresolved: Can such features persist across broader classes of systems? What role does subsystem scale play in resolving genuine dynamical critical behavior?

The present paper addresses the challenge by investigating DQPTs in a strongly interacting bosonic system, whose strong interaction characteristics permit exact solutions via mapping to spinless fermions. We focus on the subsystem Loschmidt echo (SLE)---a spatially resolved measure of dynamical fidelity---to explore the feasibility and mechanism of detecting DQPTs using local operators. Our analysis reveals that singular behavior in the SLE emerges only when the subsystem size approaches the full system scale, highlighting the fundamental limitations of local probes in capturing global dynamical criticality. More importantly, we find that low temporal resolution may obscure the true nature of the dynamics, underscoring the need for ultrahigh time resolution to distinguish genuine singularities.

The paper is organized as follows. In Section II, we present the specific model under study along with relevant parameter settings, and introduce the methods used for real-time evolution as well as the computation of related physical quantities. Section III briefly describes how the quench protocol is initiated and details the dynamical evolution for different initial states, elucidating the properties of DQPTs in the Bose-Hubbard model from both spatial and temporal perspectives. Finally, a concise summary of the paper is provided in Section IV.

\section{Model and method}

In our previous work \cite{Li2025}, we demonstrated that variations in the strength of the external harmonic potential only affect the timing of the phase transition point without causing its disappearance. Therefore, for simplicity, this work considers the simplest one-dimensional Bose-Hubbard model:
\begin{equation*}
\begin{aligned}
    {\hat H_{{\rm{B}}}} =&  - J\sum\limits_{l = 1}^{L - 1} {\hat b_l^\dag {{\hat b}_{l + 1}}}  + h.c. + \frac{U}{2}\sum\limits_{l = 1}^L {{{\hat n}_l}({{\hat n}_l} - 1)},
\end{aligned}
\end{equation*}
where $\hat b_l^\dag$ and $\hat b_l$ are the creation and annihilation operators for bosons at the lattice site $l$, and ${\hat n_l}$ is the particle number operator. $J$ denotes the tunnelling strength between nearest-neighbor sites (throughout the paper, we set $J=1$ as the energy unit, the time unit fixed as $\hbar /J$ and set $\hbar  = 1$), $U$ represents the on-site interaction strength, and $L$ is the lattice length. The total number of particles is conserved, satisfying $N = \sum\limits_{l = 1}^L {{{ n}_l}}$. $\hat b_l^\dag$ and $\hat b_l$ obey commutation relations $\left[ {{{\hat b}_j},{{\hat b}_l}} \right] \equiv {\hat b_j}{\hat b_l} - {\hat b_l}{\hat b_j} = 0$ and $\left[ {{{\hat b}_j},\hat b_l^\dag } \right] \equiv {\hat b_j}\hat b_l^\dag  - \hat b_l^\dag {\hat b_j} = {\delta _{jl}}$.

We investigate the entire dynamical process using analytical methods in the strong interaction limit ($U \to \infty $), with the additional constraint $\hat b_l^2 = \hat b_l^{\dag 2} = 0$. The system can be mapped to spinless fermions via the Jordan-Wigner transformation:
\begin{equation*}
\begin{aligned}
    {\hat b_l} = \exp \left( {{\rm{i}}\pi \sum\limits_{1 \le s < l} {\hat f_s^\dag } {{\hat f}_s}} \right){\hat f_l},
\end{aligned}
\end{equation*}
\begin{equation*}
\begin{aligned}
    \hat b_l^\dag  = \hat f_l^\dag \exp \left( { - {\rm{i}}\pi \sum\limits_{1 \le s < l} {\hat f_s^\dag } {{\hat f}_s}} \right).
\end{aligned}
\end{equation*}

By diagonalizing the Hamiltonian while neglecting the on-site interaction term, we obtain the ground state wavefunction of a spinless free fermion system \cite{Hao2012, Hao2009, Cai2011}, which can be expressed as
\begin{equation*}
\begin{aligned}
    \left| {\Psi _F^G} \right\rangle  = \prod\limits_{n = 1}^N {\sum\limits_{l = 1}^L {{P_{ln}}} } f_l^\dag |0\rangle .
\end{aligned}
\end{equation*}
The matrix $P$ can be constructed from the lowest $N$ eigenfunctions, where each column corresponds to a single particle eigenfunction. The time evolution of the selected initial state $\left| {\Psi _F^I} \right\rangle $ is essentially the evolution of the initial matrix ${P^I}$. By directly applying the time evolution operator to modify the matrix elements of ${P^I}$, the evolution result is obtained as: ${e^{ - {\rm{i}}t{{\hat H}_F}/\hbar }}{P^I} = U{e^{ - {\rm{i}}tD/\hbar }}{U^\dag }{P^I}$. Here, ${\hat H_F}$ is the Hamiltonian of the fermions after the Jordan-Wigner transformation, while $U$ and $D$ are obtained by diagonalizing ${\hat H_F}$: ${U^\dag }{H_F}U = D$.

In this work, we investigate DQPTs starting from a product state and introduce a key parameter: the SLE \cite{Kar2025}.
\begin{equation}
\begin{aligned}
    \ln {{\cal L}_M}(t) = \frac{1}{{L - M + 1}}\sum\limits_{l = 1}^{L - M + 1} {\ln } \left\langle {\prod\limits_l^{l + M - 1} {{{\hat O}_l}} } \right\rangle ,
    \label{eq1}
\end{aligned}
\end{equation}
where ${\hat O_l}$ is the projection operator for the initial state at lattice site $l$. For each selected subsystem of size $M$, we average over all contiguous subsystems. When $M=L$ (full system size), the definition is reduced to the conventional form ${{\cal L}_L}(t) = {\left| {\left\langle {{{\psi _0}}} \mathrel{\left | {\vphantom {{{\psi _0}} {\psi (t)}}} \right. \kern-\nulldelimiterspace} {{\psi (t)}} \right\rangle } \right|^2}$ (in all subsequent discussions, the subscripts $L$ and $M$ denote the full system and subsystem, respectively.). Consequently, the LE ${{\cal L}_L}(t)$ at any time $t$ is given by
\begin{equation}
\begin{aligned}
    {{\cal L}_L}\left( t \right) & = {\left| {\langle 0|{f_l}\prod\limits_{n = 1}^N {\sum\limits_{l = 1}^L {{{\left( {P_{ln}^I} \right)}^\dag }} } \prod\limits_{n = 1}^N {\sum\limits_{l = 1}^L {{P_{ln}}} } \left( t \right)f_l^\dag \left| 0 \right\rangle } \right|^2} \\
    & = {\left| {\det \left[ {{P^{I\dag }}P\left( t \right)} \right]} \right|^2}.
    \label{eq2}
\end{aligned}
\end{equation}

We explicitly present the construction of the matrix $P^I$ for product states and the computation method for ${{\cal L}_M}$ within this strong interaction computational framework. The matrix elements of $P^I$ are generated by the following formula:
\begin{equation*}
\begin{aligned}
    P_{ln}^I = \delta \left( {l,{n_l}} \right)\delta \left( {n,\sum\limits_{i = 1}^l {{n_i}} } \right).
\end{aligned}
\end{equation*}

The SLE computation requires additional operations, where for product states with single occupation the projection operators satisfy $\left\langle {{{\hat O}_l}\left( {{n_l} = 1} \right)} \right\rangle  = \left\langle {{{\hat n}_l}} \right\rangle$ and $\left\langle {{{\hat O}_l}\left( {{n_l} = 0} \right)} \right\rangle  = 1 - \left\langle {{{\hat n}_l}} \right\rangle$ (the equality signs in the above two equations indicate that, in the strong interaction limit, the operators on both sides share identical matrix representations), effectively transforming the calculation into evaluating expectation values of particle number operator products (e.g., $\left\langle {{{\hat O}_1}\left( {{n_1} = 1} \right){{\hat O}_2}\left( {{n_2} = 0} \right)} \right\rangle  = \left\langle {{{\hat n}_1}} \right\rangle  - \left\langle {{{\hat n}_1}{{\hat n}_2}} \right\rangle$) through the matrix $P$ formalism:
\begin{equation*}
\begin{aligned}
   \left\langle {\prod\limits_i^l {{{\hat n}_i}} } \right\rangle  = \det \left[ {{P_{[i:l]}}P_{[i:l]}^\dag } \right],
\end{aligned}
\end{equation*}
where the notation $[i:l]$ represents a programming syntax indicating the extraction of rows $i$ through $l$ from the matrix, where $P_{[i:l]}^\dag$ should first undergo row selection followed by conjugate transposition.

\section{Dynamics Calculation and Analysis}

In this work, we continue to employ the charge-density wave (CDW) state as the initial state and carry out the entire quench process in the strong interaction limit. Within the extended Bose-Hubbard model \cite{Cha2024}, by adjusting the strength of intersite interactions, the ground state of the system can transition from a MI to a CDW state. When both the on-site and intersite interaction strengths approach infinity, the system becomes a product state (where different particle occupation configurations correspond to different forms of the intersite interaction potential). At the initial moment, we abruptly set the intersite interaction strength to zero to initiate the quench.

\begin{figure}[htbp]
    \centering
    \subfloat{\includegraphics[width=8.5cm,height=4.1cm]{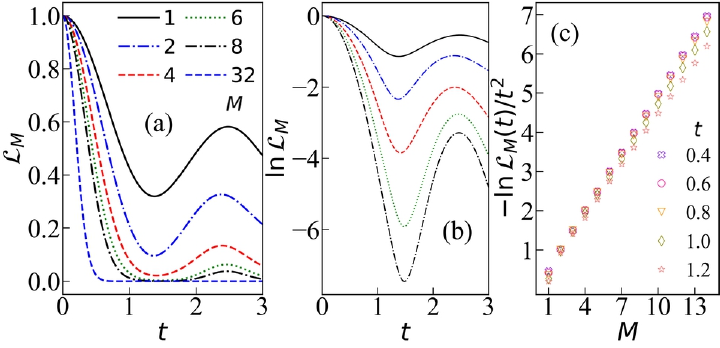}}
    \vspace{0in}
    \caption{Short-time dynamics: SLE evaluated for subsystem sizes $M = 1,2,3,4,10,32$. The  initial state is $\left| {11001100 \ldots } \right\rangle $ with $L=32$. (a) and (b) are plotted using linear and logarithmic coordinate axes, respectively. (c) Decay rate of the LE for different subsystem sizes.}
    \label{fig2}
\end{figure}

Figure \ref{fig2} shows the evolution of the SLE during short-time quench dynamics, plotted using conventional coordinates and logarithmic coordinates respectively. It can be observed that even when probing with relatively small subsystems, ${{\cal L}_M}$ decays very rapidly [Fig. \ref{fig2}(a)], which is more evident in the logarithmic plot [Fig. \ref{fig2}(b)]. Furthermore, we demonstrate that the decay of the SLE also follows an exponential law, proportional to $\exp \left( { - \lambda M{t^2}} \right)$. This behavior is detailed in Fig. \ref{fig2}(c), where $ - \ln {{\cal L}_M}/{t^2}$ calculated at different time points exhibits a clear linear dependence on the subsystem size $M$. During the early evolution stage ($t < 1$), $\lambda$ remains constant, then slightly decreases (manifested as reduced slope) near the first extremum. However, all $\ln {{\cal L}_M}$ curves exhibit well-behaved analytic properties---the functions remain smooth overall without developing cusps. This actually represents rather unfavorable news. To further investigate the properties of the SLE and its potential for detecting DQPTs, we define the subsystem rate function:
\begin{equation*}
\begin{aligned}
   {f_M}(t) =  - \frac{1}{M}\ln {{\cal L}_M}(t).
\end{aligned}
\end{equation*}

\begin{figure}[htbp]
    \centering
    \subfloat{\includegraphics[width=8.5cm,height=11.36cm]{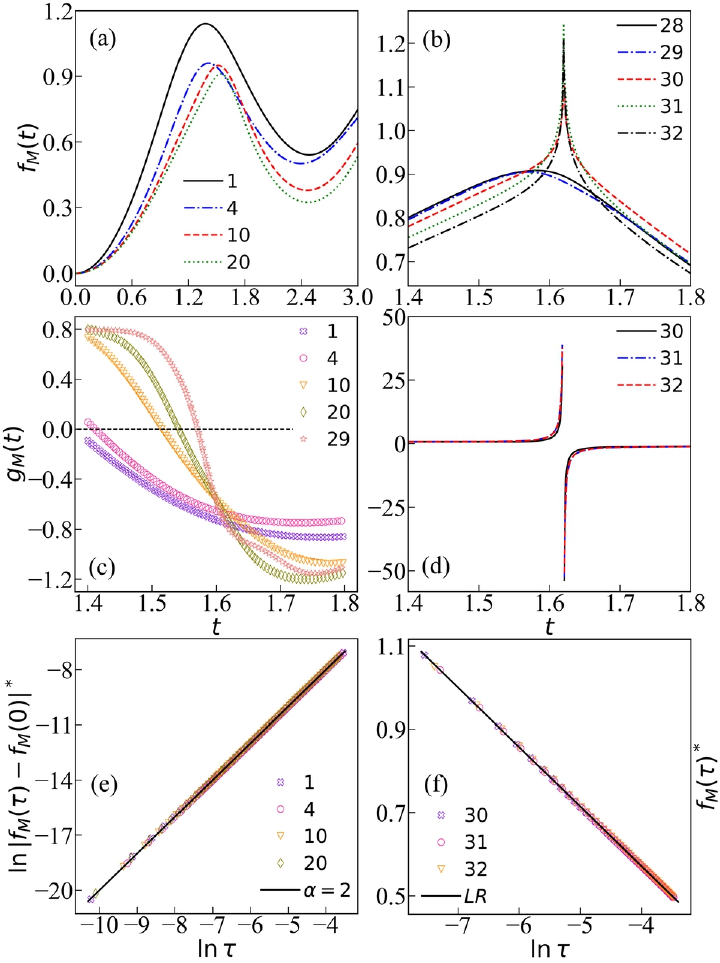}}
    \vspace{0in}
    \caption{The behavior of ${g_M}\left( t \right)$ for different subsystems near the critical point exhibits distinct characteristics. The system with size $L=32$ and initial state is $\left| {11001100 \ldots } \right\rangle $. (a) The rate function remains analytic in smaller subsystems: though peak arrival times vary across subsystems, they follow consistent trends that converge with increasing $M$. (b) Dramatic changes occur when $M$ approaches $L$: singularities emerge at $M \ge 30$, while derivative calculations confirm infinite discontinuities in ${g_M}\left( t \right)$ only when $M \ge L - 2$ (d), contrasting with continuous behavior in other subsystems (c). (e) Forcing pseudo-critical timing as $t_0$ reveals universal scaling: $\ln y$ vs $\ln x$ plots yield identical critical exponents $\alpha =2$ across subsystems near pseudo-critical points (despite varying transition locations), observable over wide size ranges (asterisks * in ordinate labels denote zero-intercept normalization). Slight deviation occurs near system scale ($M=29$, $\alpha=1.92$) where nonsingularity persists. (f) Uniform $\alpha =0$ emerges at $M \ge 30$ with single logarithmic scaling ($x$-axis only). Linear fits confirm ${f_M}\left( \tau  \right) \propto \ln \tau $ (the asterisk * indicates that both the slopes and intercepts of all data sets have been normalized).}
    \label{fig1}
\end{figure}

In a recent work, we employed the same methodology to calculate the evolution of the full system rate function in the Bose-Hubbard model, obtaining a critical exponent of $\alpha =0$ in DQPTs and demonstrating its universality across arbitrary initial product states and potential well strengths. Here, we similarly define a reduced time $\tau  = \left| (t - {t_0})/{t_0} \right|$, and for a system size of $L = 32$, further compute the derivative of the rate function with respect to time
\begin{equation*}
\begin{aligned}
   {g_M}\left( t \right) = \frac{{{\rm{d}}{f_M}\left( t \right)}}{{{\rm{d}}t}},
\end{aligned}
\end{equation*}
to investigate the critical behavior of the rate function in subsystems. Where Fig. \ref{fig1}(d) reveals ${g_M}\left( t \right)$'s divergent jumps between $ \pm \infty $ with inverse-function-like scaling, corroborating the logarithmic divergence of ${f_L}\left( t \right)$ near the critical point, though consistent critical exponents only emerge when the subsystem size $M$ approaches  $L$. When $M \le 29$ [Fig. \ref{fig1}(c)], ${g_M}\left( t \right)$ becomes continuous and crosses zero at a specific time. Close observation of the curves in Fig. \ref{fig1}(c) near their zero-crossing reveals linear behavior in all cases, implying that ${f_M}\left( \tau  \right) \propto {\tau ^2}$ behaves as a quadratic function near the zero point (which we refer to as the pseudo-critical point), with a critical exponent $\alpha =2$. Similar to Fig. \ref{fig1}(f), we present the critical behavior of the subsystem in Fig. \ref{fig1}(e), with the sole difference being that the vertical axis in Fig. \ref{fig1}(e) is also logarithmic (to eliminate the regular part of the rate function). Any smooth curve can be approximated as quadratic near an extremum, indicating the absence of phase transitions throughout the dynamical process. This firstly demonstrates that the observable selected in Eq. \ref{eq1} is unsuitable for smaller subsystems. It remains unclear whether specific observables could be identified that would yield $\alpha =0$ through measurements in small subsystems only. Comparing with results in spin models \cite{Ban2021}, we argue that a minimum subsystem size constraint exists for detecting DQPTs using subsystem operators---phase transitions become observable only when the size of the local operator exceeds this threshold. We speculate this is due to the rapid growth of the correlation length near the critical point, making it difficult for local projection operators in subsystems to capture such global dynamical behavior, which may also be an inevitable consequence of $\alpha =0$. In conclusion, the currently chosen subsystem operators do not effectively reduce the detection difficulty of DQPTs.

\begin{figure}[htbp]
    \centering
    \subfloat{\includegraphics[width=8.5cm,height=4.09cm]{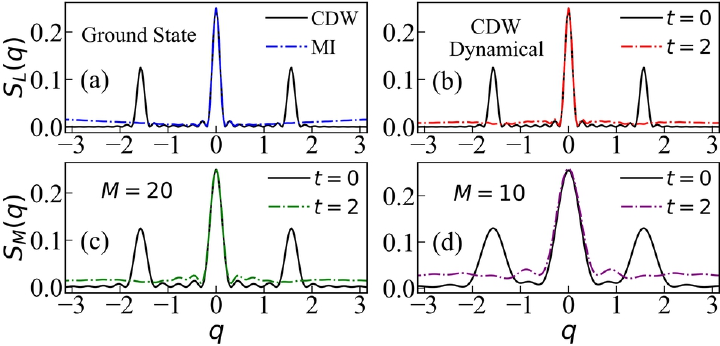}}
    \vspace{0in}
    \caption{(a) Full system structure factor ${S_L}(q)$ for different states in the ground state, with $q$ ranging over $\left[ { - \pi ,\pi } \right]$. The CDW state is $\left| {1100{\rm{1100}} \ldots } \right\rangle $ as in Fig. \ref{fig1}; the MI state is the ground state of the system Hamiltonian ${\hat H_{\rm{B}}}$ at $U/J = 100$, computed using the density matrix renormalization group (DMRG) method. (b)-(d) Structure factors of the CDW state before and after the phase transition for subsystem sizes of 32 ($L$), 20, and 10, respectively.}
    \label{fig3}
\end{figure}

While the energy gap---a key order parameter for identifying and distinguishing phases (primarily superfluid and MI phases) in ground state quantum phase transitions---becomes ineffective in dynamics due to the constraints of energy and particle number conservation, we find that the structure factor $S(q)$ as a time-dependent parameter can, to some extent, differentiate between pre- and post-DQPT phases. It should be emphasized that $S(q)$ is a continuously varying function rather than a scalar throughout the dynamical process, and thus can only provide indicative reference for the current phase of the system. Nevertheless, it not only correlates with the LE results but can also be straightforwardly extended to subsystems, defined as follows:
\begin{equation*}
\begin{aligned}
   {S_M}(q) = \frac{1}{{{M^2}}}\sum\limits_{j,k}^M {{e^{iq(j - k)}}\langle {{\hat n}_j}{{\hat n}_k}\rangle } .
\end{aligned}
\end{equation*}

As shown in Fig. \ref{fig3}(a), the CDW initial state used in our study exhibits distinct symmetric peaks at $q \ne 0$, whereas the MI state does not. Before and after the DQPTs, the structure factor of the CDW state undergoes significant changes: the original symmetric peaks disappear, and the overall distribution becomes consistent with that of the MI state [Fig. \ref{fig3}(b)]. Combined with the conclusion from Fig. \ref{fig1}(b) that the phase transition occurs around $t=1.6$, this can be interpreted as a transition from the CDW state to the MI state. The observational characteristics provided by the structure factor are more robust than those of the SLE. Even when the subsystem size $M$ is reduced to 20 [Fig. \ref{fig3}(c)] or even 10 [Fig. \ref{fig3}(d)], although the overall effect weakens, the disappearance of symmetric peaks remains clearly observable.

\begin{figure}[htbp]
    \centering
    \subfloat{\includegraphics[width=8.5cm,height=4.11cm]{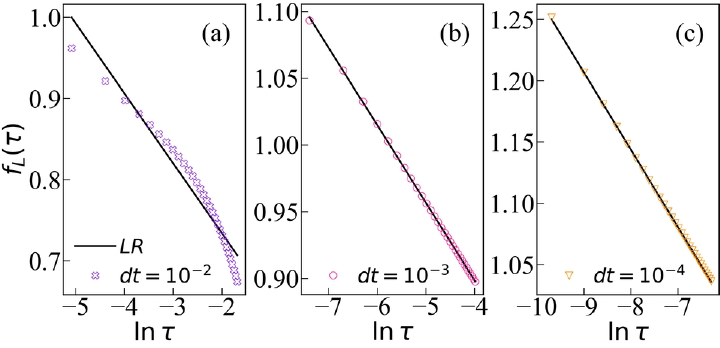}}
    \vspace{0in}
    \caption{Rate functions computed using different temporal resolutions (points) along with their linear fits (solid black lines), all fitted using the first 30 temporal data points prior to the phase transition (corresponding to time scales of ${10^{ - 1}},{10^{ - 2}},{10^{ - 3}}$, respectively). The initial state and system parameters are identical to those in Fig. \ref{fig1}. From (a) to (c), the temporal resolution progressively increases, and the computed data points approach closer to the critical point.}
    \label{fig4}
\end{figure}

On the temporal axis, both the time scale at which the phase transition occurs (specifically when the nonanalytic part of the rate function becomes dominant) and the temporal resolution required to detect DQPTs are key concerns of our investigation. A priori, the critical exponent $\alpha=0$ leads to an inevitable consequence: the logarithmic divergence progresses so gradually that one must probe extremely close to the critical point to capture the rapid growth and eventual singularity of the rate function. Insufficient temporal resolution may prevent accurate characterization of singular behavior in the rate function. Figure \ref{fig4} demonstrates significant discrepancies in linear fitting results obtained at different temporal resolutions. It is clearly observable that within the selected data points, the derivative of ${f_L}(\tau |dt = {10^{ - 2}})$ decreases as $\ln \tau $ decreases [Fig. \ref{fig4}(a)], whereas the derivatives of ${f_L}(\tau |dt = {10^{ - 3}},{10^{ - 4}})$ remain constant, exhibiting excellent linearity [Figs. \ref{fig4}(b) and \ref{fig4}(c)]. This indicates that when detecting DQPTs in the Bose-Hubbard model, inadequate temporal resolution not only hinders effective approach to the critical point and precise determination of critical exponents, but may also lead to complete missing of the critical point due to the predominance of analytic components in the data points.

\begin{figure}[htbp]
    \centering
    \subfloat{\includegraphics[width=8.5cm,height=3.98cm]{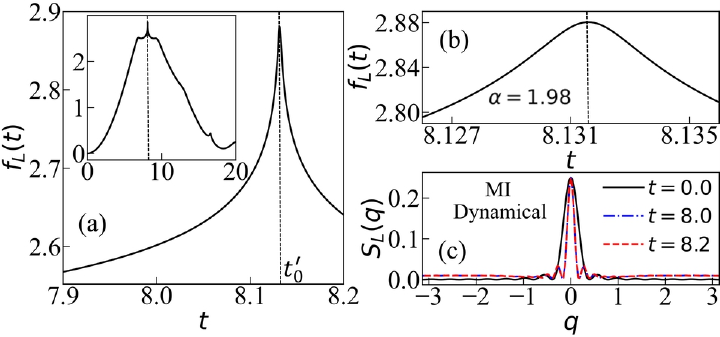}}
    \vspace{0in}
    \caption{(a)-(b) Evolution of the rate function under different time scales, with the MI state $\left| {0 \ldots 011 \ldots 110 \ldots 0} \right\rangle $ as the initial state. The time scales correspond to orders of magnitude $10^1$, $10^{-1}$, and $10^{-3}$ respectively. $t_{0}^{'}$ denotes the time at which the extremum appears (not a critical point), with the three dashed lines sharing the same $t_{0}^{'}$. (c) Structure factors of the MI state calculated at different time points, with $q$ ranging over $\left[ { - \pi ,\pi } \right]$.}
    \label{fig5}
\end{figure}

The results in Fig. \ref{fig4} suggest, to some extent, that DQPTs occur within a time scale of ${10^{ - 3}}$ to ${10^{ - 1}}$, which is further corroborated by calculating ${f_L}(t)$ for the state $\left| {0 \ldots 011 \ldots 110 \ldots 0} \right\rangle $. The structure factor results in Fig. \ref{fig5}(c) confirm that $\left| {0 \ldots 011 \ldots 110 \ldots 0} \right\rangle $ lies within the insulating regime and undergoes no phase transition under the evolution parameters selected in this work. As shown in Fig. \ref{fig5}(a), the rate function exhibits cusp-like features similar to those in Fig. \ref{fig1}(b) at time scales of ${10^{ 1}}$ or ${10^{ - 1}}$. However, when the time scale is reduced to ${10^{ - 3}}$ [Fig. \ref{fig5}(b)], the peak of the rate function appears smooth without divergence. Collectively, these results indicate that DQPTs in the Bose-Hubbard model generally manifest on a time scale of approximately ${10^{ - 2}}$, and accurate detection requires a temporal resolution of at least ${10^{ - 3}}$.

Our work not only advances theoretical understanding but also provides precise references for experimental and numerical studies of DQPTs in the Bose-Hubbard model.

\section{Conclusion}

In summary, we have investigated DQPTs in a hard-core bosonic system using the SLE as a key diagnostic tool. Our analysis reveals that: (1) The SLE exhibits abrupt singular behavior dependent on subsystem size, yielding a critical exponent of zero. However, nonanalytic behavior---the characteristic feature of DQPTs---only emerges when the subsystem dimension approaches the full system size (the local operators employed in this work require a minimum size of $L-2$), highlighting fundamental limitations in detecting DQPTs using local operators in small subsystems; (2) In the Bose-Hubba model, spurious divergences may appear even at low temporal resolution, emphasizing the necessity of ultrahigh time resolution to distinguish genuine singularities from smooth variations (acquire measurements within a time scale of ${10^{ - 2}}$ using a temporal resolution of at least ${10^{ - 3}}$.); (3) The evolution of the structure factor suggests a potential transition from a CDW state to a MI state during the quench process. Although the subsystem structure factor may not serve as the most precise parameter for characterizing DQPTs, it remains a valuable reference in dynamical studies due to its significant role in ground state phase diagrams. These findings not only elucidate the challenges in observing DQPTs in strongly interacting bosonic systems but also prompt a reevaluation of the role of locality and resolution in diagnosing nonequilibrium quantum criticality. Our results establish that, while DQPTs are a universal dynamical phenomenon, their detection via local operators requires careful consideration of both spatial and temporal scales. \\ \\

\noindent{\bf Data availability} \\
The data that support the findings of this article are openly available.

\bibliography{apssamp}% Produces the bibliography via BibTeX.

\end{document}